\documentstyle[10pt,aps,epsf]{revtex}
\begin{document}

\title{Decoherence of the Superconducting Persistent Current Qubit}
\author{Lin~Tian$^1$, L.S.~Levitov$^1$, Caspar~H.~van~der~Wal$^4$,
J.E.~Mooij$^{2,4}$, T.P.~Orlando$^2$, S.~Lloyd$^3$,
C. J. P. M. Harmans$^4$, J.J.~Mazo$^{2,5}$}

\address{$^1$Dept. of Physics, 
Center for Material Science
$\&$ Engineering,
$^2$Dept. of Electrical Engineering and 
Computer Science, 
$^3$Dept. of Mechanical Engineering, 
Massachusetts~Institute~of~Technology;  
$^4$Dept. of Applied Physics and Delft Institute
for Microelectronics and Submicron Technologies, Delft Univ.
of Technology;
$^5$Dept. de F\'{\i}sica de la Mataeria Condensada, Universidad de 
Zaragoza
}
\date{\today}
\maketitle

\begin{abstract}
Decoherence of a solid state based qubit can be caused by coupling
to microscopic degrees of freedom in the solid. We lay out a
simple theory and use it to estimate decoherence for a recently
proposed superconducting persistent current design. All considered
sources of decoherence are found to be quite weak, leading to a
high quality factor for this qubit.
\end{abstract}

\section{Introduction}

The power of quantum logic \cite{seth_universal} depends on the
degree of coherence of the qubit dynamics
\cite{exp_qubit,pc_qubit}. The so-called ``quality factor'' of the
qubit, the number of quantum operations performed during the qubit
coherence time, should be at least $10^4$ for the quantum computer
to allow for quantum error correction
\cite{quan_error_correction}. Decoherence is an especially vital
issue in solid state qubit designs, due to many kinds of low
energy excitations in the solid state environment that may couple
to qubit states and cause dephasing.

In this article we discuss and estimate {\bf some of the main
sources of} decoherence in the superconducting persistent current
qubit proposed recently \cite{pc_qubit}. The approach will be
presented in a way making it easy to generalize it to other
systems. We  emphasize those decoherence mechanisms that
illustrate this approach, and briefly summarize the results of
other mechanisms.

\begin{figure}
        \epsfysize=5cm
        \hspace{3in}\centering{\epsfbox{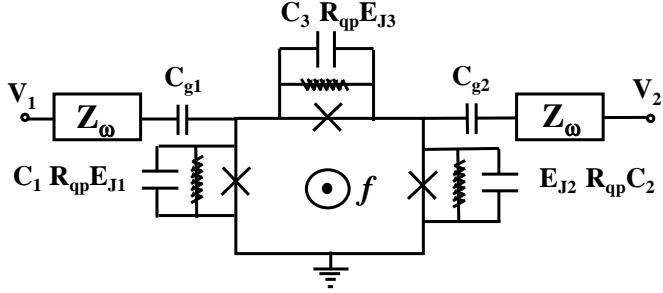}}
        \caption{
Schematic qubit design\protect\cite{pc_qubit}
consisting
of three Josephson junctions connected as shown.
Josephson energy of one of the junctions (number 3 in the figure) is adjustable
by varying the flux in the SQUID loop. The impedances $Z_\omega$ model electroma
gnetic environment
coupled to the qubit via gate capacitances $C_{g(1,2)}$. Shunt resistors model
quasiparticle subgap resistance effect. }
        \label{pc_qubit_ps}
        \end{figure}

The circuit \cite{pc_qubit} consists of three small Josephson
junctions which are connected in series, forming a loop, as shown
in Fig.~\ref{pc_qubit_ps}. The charging energy of the qubits
$E_C=e^2/2C_{1,2}$ is $\sim 100$ times smaller than the Josephson
energy $E_J=\hbar I_0/2e$, where $I_0$ is the qubit Josephson
critical current. The junctions discussed in \cite{pc_qubit} are
$200\,{\rm nm}$ by $400\,{\rm nm}$, and $E_J\approx 200\, {\rm
GHz}$.

Qubit is realized by two lowest energy states of the system
corresponding to opposite circulating currents in the loop. The
energy splitting of these states $\varepsilon_0\approx 10\, {\rm
GHz}$ is controlled by the external magnetic field flux $f$, the
barrier height is $\simeq$ 35 GHz and the tunneling amplitude
between the two states is $t\approx$ 1 GHz. The Hamiltonian
derived in \cite{pc_qubit} for the two lowest energy levels of the
qubit has the form
\begin{equation}
\begin{array}{lcl}
{\cal H}_0 & = & \left(\begin{array}{cc}
                -\varepsilon_0/2    & t(q_1, q_2) \\
                t^*(q_1, q_2) & \varepsilon_0/2  \\
                       \end{array}\right)
        \\ [3mm] \label{2by2}
\end{array}
\ ,
\end{equation}
where $t(q_1,q_2)$ is a periodic function of gate charges
$q_{1,2}$. In the tight binding approximation \cite{pc_qubit},
$t(q_1, q_2) = t_1 + t_2 e^{-i \pi q_1/e }
                        + t_2 e^{ i \pi q_2/e }$,
where $t_1$ is the amplitude of tunneling between the nearest
energy minima and $t_2$ is the tunneling between the next nearest
neighbor minima in the model \cite{pc_qubit}.  Both $t_1$ and
$t_2$ depend on the energy barrier height and width exponentially.
With the parameters of our qubit design, $t_2/t_1 < 10^{-3}$, the
effect of fluctuations of $q_{1,2}$ should be small.

Below we consider a number of decoherence effects which seem to be
most relevant for the design \cite{pc_qubit}, trying to keep the
approach general enough, so that it can be applied to other
designs.

\section{Basic approach}

We start with a Hamiltonian of a qubit coupled to environmental
degrees of freedom in the solid: ${\cal H}_{\rm total}={\cal
H}_{\rm Q}(\vec{\sigma})+{\cal H}_{\rm bath}(\{\xi_{\alpha}\})$,
where ${\cal H}_{\rm Q}={\cal H}_0+{\cal H}_{coupling}$:
  \begin{equation}\label{H-general}
{\cal H}_{\rm Q} = \frac{\hbar}{2}\,\left(\vec{\Delta}(t)+
\vec{\eta}(t)\right)\cdot\vec{\sigma} \ ,\qquad
\vec{\eta}=\sum\limits_{\alpha}\vec{A_{\alpha}}\hat\xi_{\alpha}\ ,
  \end{equation}
where $\vec{\sigma}=(\sigma_x,\sigma_y,\sigma_z)$ is the vector of
Pauli matrices acting on the qubit states, the vector
$\vec{\Delta}$ represents external control, and $\vec{\eta}$ is
noise due to coupling to the bath variables $\xi_{\alpha}$. In
(\ref{2by2}), $\Delta_z=-\varepsilon_0$,
$\Delta_x-i\Delta_y=t(q_1,q_2)$.

\par
The degrees of freedom that may decohere qubit dynamics are:
  \begin{itemize}
\item charge fluctuations in the gates coupling qubit states to
other states;
\item quasiparticles in the superconductor giving rise to subgap
resistance;
\item nuclear spins in the solid creating fluctuating magnetic
fields;
\item electromagnetic radiation causing damping of Rabi
oscillations;
\item coupling between qubits affecting operation of an individual
qubit.
  \end{itemize}
In all cases except the last one, the qubit is coupled to a
macroscopic number of degrees of freedom $N\gg1$ with about the
same strength $A_{\alpha}$ to each. In such a situation, the qubit
decoherence rate is much larger than the characteristic individual
coupling frequency $A_{\alpha}/\hbar$. This means that dephasing
happens on a shorter time scale than it would have taken to create
an entangled state of the qubit and one particular element of the
bath. In other words, on the decoherence time scale each element
of the bath remains in its initial state with probability
$1-O(1/N)$, and it is only due to a large number of relevant
degrees of freedom $N$ that the state of the qubit is
significantly affected on this time scale.
\par

This observation makes the analysis quite simple, especially
because the condition $N\gg1$ allows one to replace
generally noncommuting quantum
variables $\widehat\xi_{\alpha}(t)$ by bosonic fields
$\eta_{x,y,z}(t)$ fluctuating in
time. (Because at large $N$ the commutators
$[\eta_i(t),\eta_j(t)]$
are well approximated by $c-$numbers.)
As a result, the problem becomes equivalent to
that of longitudinal and transverse spin relaxation times $T_1$
and $T_2$ in NMR, corresponding to the noise $\eta_i(t)$ either
flipping the qubit spin, or contributing a random phase to the
qubit states evolution, respectively. Thus we can use the standard
Debye--Bloch theory of relaxation in two-level systems.

To adapt this theory to our problem, we assume, without loss of
generality, that $\vec{\Delta}(t)\parallel\hat{z}$ and is constant
as a function of time. Then one can eliminate the term
$\frac{1}{2}\vec{\Delta}\cdot\vec{\sigma}$ by going to the frame
rotating around the $z-$axis with the Larmor frequency
$\Delta=|\vec{\Delta}|$. In the rotating frame the Hamiltonian
(\ref{H-general}) becomes:
  \begin{equation}\label{H-rotated}
\widetilde {\cal H}_{\rm Q} = \frac{\hbar}{2}\,\left(
\eta_{\parallel}(t)\sigma_z+e^{-i\Delta t}\eta_{\perp}(t)\sigma_+
+e^{i\Delta t}\eta^\ast_{\perp}(t)\sigma_-\right) \ ,
  \end{equation}
where $\eta_{\parallel}(t)$ and $\eta_{\perp}(t)$ correspond to
components of vector $\vec\eta(t)$ in (\ref{H-general}) parallel
and perpendicular to $\vec\Delta$, respectively.

The time evolution due to noise $\vec\eta(t)$ is given by the
evolution operator $T\exp\left(-i\int\widetilde {\cal H}_{\rm
Q}(t')dt'\right)$ written in the rotating Larmor basis. However,
for a simple estimate below we ignore noncommutativity of
different parts of the Hamiltonian (\ref{H-rotated}), and consider
a c-number phase factor instead of an operator exponent.

Then the decoherence can be characterized using the function
  \begin{equation}\label{R(t)}
R(t)={\rm max}\,\left[\langle\phi^2_{\parallel}(t)
\rangle,\,\langle|\phi_{\perp}(t )|^2\rangle\right] \ ,
  \end{equation}
where $\langle...\rangle$ stands for ensemble average, and
  \begin{equation}
\phi_{\parallel}(t)=\int\limits^t_0\eta_{\parallel}(t')dt' \
,\qquad \phi_{\perp}(t)= \int\limits^t_0
e^{-i\Delta t'}\eta_{\perp}(t')dt' \nonumber
  \end{equation}
The function $R(t)$ grows with time, and one can take as a measure
of decoherence the time $\tau$ for which $R(\tau)\simeq 1$.
There are several assumptions implicit in this criterion.
First, we ignore noncommutativity of different terms in
(\ref{H-rotated}), which is legitimate at short times, when $R(t)\ll1$.
Second, we include in (\ref{R(t)}) the
zero-point fluctuations of $\widehat\xi_{\alpha}(t)$. The issue
of decoherence due to zero-point motion in some cases
can be subtle. However, since
including the zero-point fluctuations in $R(t)$ can only
overestimate the rate of loosing coherence, one expects
the criterion $R(\tau)\simeq 1$ to still give a good lower bound
on decoherence time.

Finally, we note that (\ref{R(t)}) contains
statistical average over an ensemble of bath realizations.
Hence care needs to be taken in the interpretation of $\tau$ when the
bath is ``frozen''  into a particular  configuration so that the
ensemble averaging does not apply. In this situation one has
to distinguish between decohering individual qubit
dynamics and averaged dynamics of a qubit array.
An example of such a situation is provided by the problem
of coupling to the nuclear spins, a system with long relaxation times.

Since $\vec\eta=\sum\limits_{\alpha}\vec
A_{\alpha}\hat\xi_{\alpha}(t)$, it is the time evolution of
${\widehat\xi}_{\alpha}(t)$ defined by ${\cal H}_{\rm bath}$ that
is what eventually leads to decoherence. One can express
quantities of interest in terms of the noise spectrum of the
components of $\vec\eta$:
  \begin{eqnarray}\label{parallel}
\langle\phi^2_{\parallel}(t)\rangle &=&\int d\omega
\frac{|1-e^{i\omega t}|^2}{2\pi\omega^2}
\langle\eta_{\parallel}(-\omega)\eta_{\parallel}(\omega)\rangle
\\
\label{perp} \langle|\phi_{\perp}(t)|^2\rangle &=&\int d\omega
\frac{|1-e^{i\omega t}|^2}{2\pi\omega^2}
\langle\eta_{\perp}(-\omega-\Delta)\eta_{\perp}(\omega+\Delta)\rangle
  \end{eqnarray}
In thermal equilibrium, by virtue of the Fluctuation--Dissipation
theorem, the noise spectrum in the RHS of (\ref{parallel}) and
(\ref{perp}) can be expressed in terms of the out-of-phase part of
an appropriate susceptibility.

\section{Estimates for particular mechanisms}

Here we discuss the above listed decoherence mechanisms and use
the expressions (\ref{parallel}) and (\ref{perp}) to estimate the
corresponding decoherence times. We start with the effect of {\bf
charge fluctuations on the gates} due to electromagnetic coupling
to the  environment modeled by an external impedance $Z_\omega$
(see Fig.~\ref{pc_qubit_ps}), taken below to be of order of 400
$\Omega$, the vacuum impedance.

The dependence of the qubit Hamiltonian on the gate charges
$q_{1,2}$ is given by (\ref{2by2}), where $q_{1,2}$ vary in time
in response to the fluctuations of gate voltages, $\delta
q_{1,2}\approx C_g \delta V_{g(1,2)}$, where the gate capacitance
is much smaller than the junction capacitance: $C_g\ll C_{1,2}$.
The gate voltage fluctuations are given by the Nyquist formula:
$\langle\delta V_g(-\omega)\delta
V_g(\omega)\rangle=2Z_\omega\hbar\omega\coth\hbar\omega/kT$.

In our design, $|t(q_1,q_2)|\ll\varepsilon_0$, and therefore
fluctuations of $q_{1,2}$ generate primarily transverse noise
$\eta_{\perp}$ in (\ref{H-rotated}), $\eta_{\perp}(t)\simeq
(2\pi/\hbar e)t_2C_g\delta V_g(t)$. In this case, according to
(\ref{perp}), we are interested in the noise spectrum of $\delta
V_g$ shifted by the Larmor frequency $\Delta$. Our typical $\Delta
\simeq 10\, {\rm GHz}$ is much larger than the temperature
$k_BT/h= 1\, {\rm GHz}$ at $T=50\,{\rm mK}$, and thus one has
$\omega\simeq\Delta\gg kT/\hbar$ in the Nyquist formula.

The Nyquist spectrum is very broad compared to Larmor frequency
and other relevant frequency scales, and thus in (\ref{perp}) we
can just use the $\omega=\Delta$ value of the noise power.
Evaluating 
$\int |(1-e^{i\omega t})/\omega|^2d\omega
=2\pi t$, we obtain
  \begin{equation}
R(t)=\langle|\phi_{\perp}(t)|^2\rangle
=\frac{2t}{\hbar}\left(\frac{2\pi}{e}t_2C_g\right)^2\,\Delta
Z_{\omega=\Delta}
  \end{equation}
Rewriting this expression as $R(t)=t/\tau$, we estimate the
decoherence time as
  \begin{equation}
\tau=\Delta^{-1}\frac{\hbar}{2e^2}Z^{-1}_{\omega=\Delta}
\left(\frac{e^2}{2\pi C_g t_ 2}\right)^2
  \end{equation}
where $\hbar/2e^2\simeq 4\, {\rm k\Omega}$. In the qubit design
$e^2/2C_g\simeq 100\, {\rm GHz}$, and $t_2\simeq 1\, {\rm MHz}$
when $t_2/t_1\le 10^{-3}$. With these numbers, one has $\tau=0.1\,
{\rm s}$.

The next effect we consider is dephasing due to {\bf
quasiparticles on superconducting islands}. At finite temperature,
quasiparticles are thermally activated above the superconducting
gap $\Delta_0$, and their density is $\sim\exp(-\Delta_0/kT)$. The
contribution of quasiparticles to the Josephson junction dynamics
can be modeled as a shunt resistor, as shown in
Fig.~\ref{pc_qubit_ps}. The corresponding {\it subgap resistance} is
inversely proportional to the quasiparticle density, and thus
increases exponentially at small temperatures: $R_{\rm qp} \approx
R_n \exp\Delta_0/kT$, where $R_n$ is the normal state resistance
of the junction. For Josephson current $I_0=0.2 \ \mu{\rm A}$,
$R_n \approx 1.3\, {\rm k} \Omega$. At low temperatures the subgap
resistance is quite high, and thus difficult to measure
\cite{subgap_exp}. For estimates below we take $R_{\rm qp} =
10^{11} \ \Omega$ which is much smaller than what follows from the
exponential dependence for $T=50\, {\rm mK}$.

The main effect of the subgap resistance in the shunt resistor
model is generating normal current fluctuations which couple to
the phase on the junction. The Hamiltonian describing this effect
is
  \begin{equation}\label{H-subgap}
{\cal H}^{\rm qp}_{\rm
coupling}=\sum\limits_{i}\frac{\hbar}{2e}\varphi_i I^{\rm
qp}_i(t)\ ,
  \end{equation}
where $i\!=\!1,2,3$ labels Josephson junctions. Projecting
(\ref{H-subgap}) to the two qubit states, one obtains the
Hamiltonian (\ref{H-general}) with $\eta_z(t)=I^{\rm qp}_i(t)/e$,
$\eta_{x,y}=0$.

The noise spectrum of the quasiparticle current is given by
Nyquist formula:
  \begin{equation}
\langle I^{\rm qp}(-\omega)I^{\rm qp}(\omega)\rangle =2R_{\rm
qp}^{-1}\hbar\omega\coth(\hbar\omega/kT)
  \end{equation}
After rotating the basis and transforming the problem to the form
(\ref{H-rotated}) we have $\eta_{\perp}(t)\simeq
(t_1/\varepsilon_0)\eta_{\parallel}(t)$, where
$\eta_{\parallel}(t)\simeq I^{\rm qp}_i(t)/e$ since $t_1\ll
\varepsilon_0$.

The analysis of $\langle|\phi_{\perp}(t)|^2\rangle$ and
$\langle|\phi_{\parallel}(t)|^2\rangle$ is similar to that
described above for charge fluctuations on the gates, and one
obtains \hspace{0.3cm} $R_{\perp}(t) = 2t(t_1/\varepsilon_0)^2
\hbar\Delta/(e^2R_{\rm qp})$, and
$R_{\parallel}(t)=2t\,kT/(e^2R_{\rm qp})$ which gives
  \begin{equation}
\tau=\,{\rm min}\,\left[\tau_{\perp}, \tau_{\parallel}\right]
=\,{\rm min}\,\left[ \frac{e^2R_{\rm
qp}}{2\hbar\Delta}\left(\frac{\varepsilon_0}{t_1}\right)^2,\
\frac{e^2R_{\rm qp}}{2kT}\right]
  \end{equation}
Taking $R_{qp}=10^{11}\,\Omega$, $T=50\, {\rm mK}$, and
$\varepsilon_0/t_1=100$, the decoherence times are
$\tau_{\parallel}=1\, {\rm ms}$ and $\tau_{\perp}=10\, {\rm ms}$.

The decoherence effect of {\bf nuclear spins} on the qubit is due
to their magnetic field flux coupling to the qubit inductance.
Alternatively, this coupling can be viewed as Zeeman energy of
nuclear spins in the magnetic field $\vec B(r)$ due to the qubit.
The two states of the qubit have opposite currents, and produce
magnetic field of opposite sign. The corresponding term in
(\ref{H-general}) is
  \begin{equation}\label{H-nuclei}
{\cal H}_{\rm coupling}= -\sigma_z\sum\limits_{r=r_i} \mu\vec
B(r)\cdot\vec{\widehat s}(r)
  \end{equation}
where $r_i$ are positions of nuclei, $\mu$ is nuclear magnetic
moment and ${\widehat s}(r_i)$ are spin operators.

Nuclei are in thermal equilibrium, and their spin fluctuations can
be related to the longitudinal relaxation time $T_1$ by the
Fluctuation-Dissipation theorem. Assuming that different spins are
uncorrelated, one has
  \begin{equation}\label{spin-noise}
\langle s_\omega(r) s_{-\omega}(r)\rangle
=2k_BT\frac{\chi''(\omega)}{\omega} =\frac{2k_B
T_1\chi_0}{1+\omega^2T^2_1} \ ,
  \end{equation}
where $\chi_0=1/k_BT$ is static spin susceptibility.

The spectrum (\ref{spin-noise}) has a very narrow width set by the
long relaxation time $T_1$. This width is much less then $k_BT$
and $\Delta$. As a result, only longitudinal fluctuations
$\eta_{\parallel}$ survive in (\ref{parallel}) and (\ref{perp}).
One has
  \begin{equation}\label{phi2}
\langle\phi^2_{\parallel}(t)\rangle = \int
d\omega\frac{|1-e^{i\omega t}|^2}{2\pi\hbar^2\omega^2}
\sum\limits_{r=r_i}\mu^2 B^2(r)\langle s_\omega(r)
s_{-\omega}(r)\rangle \ .
  \end{equation}
Plugging the spectrum (\ref{spin-noise}) in (\ref{phi2}) and
integrating, one obtains
  \begin{eqnarray}
R(t)=\frac{T_1}{\tau^2_0}\left(|t|-T_1+T_1e^{-|t|/T_1}\right) ,
\nonumber\\ \\ \tau_0=\left(\int\frac{2\mu^2}{\hbar^2}n(r)B^2(r)
d^3r\right)^{-1/2} , \nonumber \label{t/tau}
  \end{eqnarray}
where $n(r)$ is the nuclei concentration. The {\bf
ensembled-averaged} decoherence time  that  defined by
$R(\tau)\simeq 1$ is then estimated as:
  \begin{equation}\label{tau}
\tau =\cases{\tau_0 \qquad\quad \mbox{for} \quad T_1>\tau_0 \cr
\tau^2_0/T_1 \quad \mbox{   for} \quad T_1<\tau_0}
  \end{equation}
In superconducting Al, nuclear spin relaxation time is strongly
varying with temperature: $T_1\simeq (300/T\, [{\rm K}])
e^{\Delta/k_BT}\,{\rm s}$. At $T=50\,{\rm mK}$, the time $T_1$ is
of order of minutes, which exceeds all time scales relevant for
qubit operation. To estimate $\tau_0$, we use the magneton
$\mu\simeq e\hbar/Mc$, where $M$ is proton mass, and $\int
B^2(r)d^3r\simeq 10^{-5}\Phi^2_0/w$, where $w\simeq 0.5\, \mu{\rm
m}$ is the thickness of Al wires in the circuit, and
$\Phi_0=hc/2e$ is the flux quantum. The resulting $\tau_0\simeq 3
\times 10^{-8}\,{\rm s}\ll T_1$.

According to (\ref{tau}), one apparently obtains a worryingly
short time $\tau=\tau_0$. However, we note that this result
corresponds to ensemble averaging, and one should be careful in
applying it to an individual qubit.

The physical picture is that the nuclei spin configuration stays
the same over times $\le T_1$. At such times the perturbation
$\vec\eta$ due to spins has essentially no time dependence, and so
nuclei can be viewed as sources of random {\it static} magnetic
field. The fluxes of this field induced on the  qubits depend on
initial conditions, and are uncorrelated for different qubits.
Typical value of this flux corresponds to the change in Larmor
frequency of order of $\delta\Delta\simeq\tau^{-1}_0\simeq 30 \,
{\rm MHz}$.

To summarize, for an individual qubit the effect of nuclear spins
is equivalent to a random detuning caused by random change in
$\Delta$. For an ensemble of qubits, there will be a distribution
of Larmor frequencies of width $\delta\Delta\simeq 30 \, {\rm
MHz}$, even if all qubits are identical. However, since the qubit
phase can be kept coherent within a time $\le T_1$, an indirect
observation of Rabi oscillations is still possible by using the
so-called ``spin-echo technique.''

A similar theory can be employed to estimate the effect due to
magnetic impurities. The main difference is that for impurity
spins the relaxation time $T_1$ is typically much shorter than for
nuclear spins. If $T_1$ becomes comparable to the qubit operation
time, the ensemble averaged quantities will describe a {\it real}
dephasing of an individual qubit, rather than effects of
inhomogeneous broadening, like for nuclear spins.

\section{Other Mechanisms}

Some sources of decoherence are not amenable to the basic approach
considered above, such as radiation losses which we estimate to
have $\tau\simeq 10^3\,{\rm s}$.

Another such source of decoherence is caused by the  magnetic
dipole interaction between the qubits. This {\bf interaction
between qubits} is described by
  \begin{equation}\label{qubit-qubit}
{\cal H}_{\rm
coupling}=\sum\limits_{i,j}\hbar\lambda_{ij}\sigma^{(i)}_z\otimes\sigma^{(j)}_z
\ ,\qquad \hbar\lambda_{ij}\approx\frac{\mu_i\mu_j}{|r_i-r_j|^3}
  \end{equation}
This interaction is strongest for nearest neighbors. For a square
lattice of qubits with the spacing $R=10\, \mu{\rm m}$, one has
the nearest neighbor coupling $\lambda\simeq 6\,{\rm kHz}$. The
corresponding decoherence time $\tau=\lambda^{-1}\simeq 0.2\,{\rm
ms}$ is relatively short.

Several alterations of the design can be implemented to reduce the
effect of qubit-qubit interaction. One can arrange qubits in pairs
with opposite sign of circulating currents. This will eliminate
dipole moment of a pair, and reduce coupling between different
pairs to a somewhat weaker quadrupole interaction. The same result
can be achieved by using a superconducting base plane, in which
magnetic dipoles will be imaged by dipoles of opposite sign, which
will partially cancel the qubit-qubit coupling. Also, one can
detune Larmor frequencies of neighboring qubits, moving them apart
by more than $\lambda$, which will make couplings
(\ref{qubit-qubit}) off-resonant and reduce their effect.

Unwanted coupling between qubits is a common problem in quantum
computers. Sophisticated decoupling techniques that have been
developed for NMR designs \cite{Cory}, could equally be used here.
The idea is to apply a sequence of single bit operations that
effectively average out the coupling Hamiltonian over time. Such
methods would also be effective for reducing the coupling to the
environment \cite{Viola}. These techniques are fully compatible
with quantum computation and could be used to lengthen
significantly the effective coherence times.

\section{Summary}

Our analysis shows that for the qubit design \cite{pc_qubit} the
decoherence time is limited by qubit-qubit coupling. By using
methods discussed above the decoherence time can be made at least
$1\, {\rm ms}$ which for $f_{\rm Rabi}=100\,{\rm MHz}$ gives a
quality factor of $10^5$, passing the criterion for quantum error
correction.

In addition to the effects we discussed, some other decoherence
sources are worth attention, such as low frequency charge
fluctuations resulting from electron hopping on impurities in the
semiconductor and charge configuration switching near the gates
\cite{charge_effect_exp}. These effects cause $1/f$ noise in
electron transport, and may contribute to decoherence at low
frequencies. Also, we left out the effect of the {\it ac} field
coupling the two low energy states of the qubit to higher energy
states. Results of our numerical simulations of the coupling
matrix in the qubit \cite{pc_qubit} show that Rabi oscillations
can be observed even in the presence of the {\it ac} excitation
mixing the states (to be published elsewhere).

\acknowledgements \noindent This work is supported by ARO grant
DAAG55-98-1-0369, NSF Award 67436000IRG, Stichting voor
Fundamenteel Onderzoek der Materie and the New Energy and
Industrial Technology Development Organization.

\end{document}